\documentclass[conference]{IEEEtran}
\IEEEoverridecommandlockouts
\usepackage{cite}
\usepackage{amsmath,amssymb,amsfonts}
\usepackage{algorithmic}
\usepackage{graphicx}
\usepackage{textcomp}
\usepackage[dvipsnames]{xcolor}
\usepackage{t1enc}
\usepackage[latin2]{inputenc}
\usepackage{qcircuit}
\usepackage{bbm}
\usepackage{cite} 

\usepackage{pgf}
\usepackage{array, makecell}
\usepackage[caption=false]{subfig}
\usepackage[normalem]{ulem}

\DeclareMathOperator*{\argmin}{arg\,min}
\newcommand{\R}{\ensuremath{\mathbb{R}}}
\DeclareMathOperator*{\minimize}{minimize\thickspace}

\def\BibTeX{{\rm B\kern-.05em{\sc i\kern-.025em b}\kern-.08em
    T\kern-.1667em\lower.7ex\hbox{E}\kern-.125emX}}
    
\begin{document}

\makeatletter
    \newcommand{\linebreakand}{%
      \end{@IEEEauthorhalign}
      \hfill\mbox{}\par
      \mbox{}\hfill\begin{@IEEEauthorhalign}
    }
    \makeatother

\title{Quantum Optimization for the Graph Coloring Problem with Space-Efficient Embedding}

\author{
\linebreakand
\IEEEauthorblockN{Zsolt Tabi}
\IEEEauthorblockA{\textit{Ericsson Hungary and}\\
\textit{E\"otv\"os Lor\'and University} \\
Budapest, Hungary \\
zsolt.tabi@ericsson.com}
\and
\IEEEauthorblockN{Kareem H. El-Safty}
\IEEEauthorblockA{\textit{Wigner Research Centre for Physics} \\
Budapest, Hungary \\
kareem.elsafty@wigner.hu}
\and
\IEEEauthorblockN{Zs\'ofia Kallus}
\IEEEauthorblockA{\textit{Ericsson Research} \\
Budapest, Hungary \\
zsofia.kallus@ericsson.com}
\and 
\IEEEauthorblockN{P\'eter H\'aga}
\IEEEauthorblockA{\textit{Ericsson Research} \\
Budapest, Hungary \\
peter.haga@ericsson.com}\\
\linebreakand
\IEEEauthorblockN{Tam\'as Kozsik}
\IEEEauthorblockA{\textit{Faculty of Informatics}\\
\textit{E\"otv\"os Lor\'and University} \\
Budapest, Hungary \\
kto@elte.hu}
\and
\IEEEauthorblockN{Adam Glos}
\IEEEauthorblockA{\textit{Institute of Theoretical}\\ \textit{ and Applied Informatics} \\
\textit{Polish Academy of Sciences}\\
Gliwice, Poland \\
aglos@iitis.pl}
\and 
\IEEEauthorblockN{Zolt\'an Zimbor\'as}
\IEEEauthorblockA{\textit{Wigner Research Centre for Physics and} \\
\textit{MTA-BME Lend\"ulet QIT Research Group} \\
\textit{Budapest University of Technology}\\
Budapest, Hungary \\
zimboras.zoltan@wigner.hu}
}

\maketitle

\begin{abstract}
Current quantum computing devices have different strengths and weaknesses depending on their architectures. This means that flexible approaches to circuit design are necessary. We address this task by introducing a novel space-efficient quantum optimization algorithm for the graph coloring problem. Our circuits are deeper than the ones of the standard approach. However, the number of required qubits is exponentially reduced in the number of colors. We present extensive numerical simulations demonstrating the performance of our approach.  Furthermore, to explore currently available alternatives,
we also perform a study of random graph coloring on a quantum annealer to test the limiting factors of that approach, too.

\end{abstract}


\section{Introduction}

Quantum computers are expected to offer speedups over classical computers in solving various computational tasks. 
The recent demonstration of quantum computational advantage by Google researchers \cite{arute2019quantum} further strengthened the case for the practical potential of quantum computation. However, despite the many promising results, the limitations of the near-term Noisy Intermediate-Scale Quantum (NISQ) devices has also been highlighted \cite{preskill2018quantum}. Currently, various architectures and physical realizations are being considered for quantum processors, e.g., superconducting qubits \cite{corcoles2015demonstration, barends2014superconducting, ofek2016extending}, ion-trap-based systems \cite{debnath2016demonstration,monz2016realization}, integrated quantum optics \cite{qiang2018large,elshaari2020hybrid}. 

The different quantum hardware implementations have different strengths and weaknesses. For example, scaling up the number of qubits in superconducting architectures is easier than in ion trap
systems, while the latter gives rise to deeper circuits. Due to these different features, there has been an intensive discussion about how to define a suitable metric to quantify a quantum processor's performance, one measure being the so-called quantum volume \cite{cross2019validating}. Consequently, the challenge to compile large problems into programs (circuits) that minimize the number of qubits and/or the circuit depth has become of central importance in the quantum computing community.
Our paper addresses this challenge by introducing a new space-efficient embedding of the graph coloring problem. By applying this method as an input for the Quantum Approximate Optimization Algorithm \cite{farhi2014quantum}, we obtain a deeper circuit but the number of required qubits (circuit width) is exponentially reduced in the number of colors compared to the standard quadratic binary embedding method. Our numerical studies also indicates that the increase of the depth might not be that significant for larger system sizes, as one needs less levels in the space-efficient embedded version. Moreover, the number of iterations to reach nearly optimal parameters  is also significantly decreased compared to the standard version.

The paper is organized as follows. In the next section, graph coloring as a QUBO problem is reviewed. In Sec.~\ref{sec:space-efficient-method}, we present our novel space-efficient embedding method for the graph coloring problem. In order to test the current quantum hardware's performance on graph coloring, Sec.~\ref{sec:annealer} is devoted to experimenting on D-Wave's quantum annealer, covering a wide range of \emph{Erd\H{o}s-R\'{e}nyi} random graph instances. In Sec.~\ref{sec:QAOA}, we outline the Quantum Approximate Optimization Algorithm (QAOA). We also present numerical simulation results for both the standard and the space-efficient QAOA methods applied to graph coloring problems of different graphs. Finally, Sec.~\ref{sec:conclusion} summarizes our findings.

\section{Graph Coloring and its Reformulation as a QUBO problem}
\label{sec:gc_as_qubo}

In this section, we shortly review the basics of the coloring problem and its standard formulation as a Quadratic Unconstrained Binary Optimization (QUBO) problem,  we introduce the relevant
 notations and  relate our results to previous studies.

\subsection{Graph Coloring}

Graph coloring is a way of labeling the vertices of a graph with colors
%indices from a finite set, the ``colors'',  
such that no two adjacent vertices are assigned the same color. A coloring using at most $k$ different labels is called a  $k$-coloring, and the smallest number of colors needed to color a graph G is called its {\it chromatic number}.

Graph coloring has many applications in a wide range of industrial and scientific fields, such as social networks \cite{rossi2014coloring}, telecommunication \cite{bandh2009graph}, and compiler theory \cite{chaitin1982register}. Due to this, it has been in the focus of attention of researchers in computer science and operations research. Although graph coloring, in general, is NP-hard \cite{gary1979computers}, the hardness of a coloring problem is highly dependent on the graph structure and number of colors, and for  some special cases, there exists polynomial time exact solvers. For the general case, approximate solution can be achieved by using heuristic algorithms such as Tabu search \cite{hertz1987using} and Simulated Annealing \cite{johnson1991optimization}.

\subsection{Graph Coloring as a QUBO Problem}
\label{sec:qubo-problem}

QUBO is a standard model in optimization theory that is frequently used in quantum computing as
it can serve as an input for algorithms like the Quantum Approximate Optimization Algorithm (QAOA) \cite{farhi2014quantum} or Quantum Annealing (QA) \cite{kadowaki1998quantum, farhi2002quantum}. The general form of QUBO problems is the minimization of $f : \left\{ 0,1 \right\} ^N \to \R$, the pseudo-Boolean objective function of the following form:
\begin{align}
  \minimize f(x) &= x^T Q x =\sum_{i,j=1}^N Q_{ij} x_ix_j\enspace , \nonumber \\ 
  x^* &= \argmin_{x \in \{ 0,1 \}^N} f(x),  \nonumber
\end{align}
\noindent
where $Q$ is a real symmetric matrix, $f$ is often called the {\it cost function} and $x^*$ is referred to as a {\it solution bit string} or a {\it global minimizer} of $f$.

%optimisation variables, (called decision variables) representing a sequence of \textit{true} or \textit{false} values.
Such a QUBO problem is equivalent to finding the ground-state energy and  configurations of the following $N$-qubit Ising Hamiltonian \cite{lucas2014ising}:
\begin{align}
  H&= \sum_{i,j=1}^N Q_{i j} (\mathbbm{1}-Z_i) (\mathbbm{1}-Z_j), %\nonumber
 % \\
 % &= \frac{1}{2} \sum_{i < j =1}^N Q_{i j} Z_i Z_j - 2 \sum_{i=1}^N \left( \sum_{j=1}^N Q_{ij}\right)Z_i \enspace
\end{align}
\noindent
where $Z_k$ denotes the operator that acts as the Pauli-$Z$ gate on the $k$th qubit and as identity on the other qubits.

The coloring problem, similarly to several other families of NP-complete problems, can be naturally formulated as a QUBO problem. The QUBO description of the $k$-coloring problem for a graph with $n$ nodes uses $N= n {\cdot }k$ number of bits. The bits $x_{v,i}$ in this formulation have double labels $(v,i)$, where $v \in \{1, \ldots, n \}$ labels the vertices and $i \in \{1, \ldots , k \}$ labels the colors. One uses a one-hot encoding, i.e.,  if vertex $v$ is assigned the color $j$ we set $x_{v,j}=1$ and for all $i \ne j$ we set $x_{v, i}=0$. To ensure that the solution of the QUBO will satisfy such a one-hot encoding requirement, one employs a penalty term for each vertex $v$ of the form $(1 - \sum_{i=1}^k x_{v,i})^2$. Next, for all pairs $(v,w)$ of neighboring sites, one penalizes the same-color assignments by the term $\sum_{i=1}^k x_{v,i}x_{w,i} $. Thus, in total, the cost function for the $k$-coloring of a graph with $n$ nodes and adjacency matrix $A$ can be written as follows: 
\begin{equation}
    f(x) {=} C\sum_{v=1}^n \left(1 {-} \sum_{i=1}^k x_{v,i} \right)^2 {+} D\sum_{v,w=1}^n \sum_{i=1}^k A_{vw}  x_{v,i}x_{w,i},
\end{equation}
where $C$ and $D$ can be arbitrary positive numbers. The corresponding Ising model is thus: 
\begin{align}
H & = C \sum_{v=1}^n \Big(2\mathbbm{1}- \sum_{i=1}^k (\mathbbm{1}{-}Z_{v,i}) \Big)^2  \nonumber \\ 
&+   D \sum_{v,w=1}^n \sum_{i=1}^k A_{vw}(\mathbbm{1} -Z_{v,i})(\mathbbm{1}{-}Z_{w,i}). \label{eq:color_ising1}
\end{align}

\section{Space-Efficient Graph Coloring Embedding}
\label{sec:space-efficient-method}
%\hp{The graph coloring problem is usually addressed as a QUBO blabl a...}
%\cite{bergholm2018pennylane}

We now introduce a method to map the $k$-coloring problem to the ground-state problem of a Hamiltonian using only $n \lceil \log k \rceil$ instead of $n k$ qubits that are required by the standard QUBO method. This embedding will be used to set up a space-efficient QAOA method for coloring in Section~\ref{sec:QAOA}.

We will first describe the embedding of the $4$-coloring problem of an $n$-vertex graph into a $2n$-bit Hamiltonian optimization problem.
The four colors will be encoded by 2 bits ($00$, $01$, $10$, $11$). To each vertex $v$, we assign two bits $b_{v,1}$ and $b_{v,2}$, and the bit-string $(b_{v,1},b_{v,2})$ encodes the color that we assign to vertex $v$. To ensure that two neighboring vertices do not have the same color, we introduce the penalty term
\begin{align}
    &b_{v,1}b_{w,1}b_{v,2}b_{w,2} {+} (1{-}b_{v,1})(1{-}b_{w,1})(1{-}b_{v,2})(1{-}b_{w,2})\nonumber \\ &+(1{-}b_{v,1})(1{-}b_{w,1})b_{v,2}b_{w,2} 
     +b_{v,1}b_{w,1}(b_{v,2}{-}1)(b_{w,2}{-}1), \nonumber %\nonumber 
    %\\
    %&+ (1{-}b_{v,1})(1-b_{v,2})(1{-}b_{w,1})(1{-}b_{w,2}) ,
\end{align}
since this term is only zero if the colors assigned to $v$ and $w$ differ. Thus, in the case of a graph with adjacency matrix $A$, the $4$-coloring problem translates to the ground-state problem of the Hamiltonian

% \tz{the first 1+Z_w,2 should be 1-Z_w,2}
\begin{align}
H {=} \sum_{v, w=1}^n &A_{vw}   \Big((\mathbbm{1} {-} Z_{v, 1}) (\mathbbm{1} {-} Z_{v, 2})  (\mathbbm{1} {-} Z_{w, 1}) (\mathbbm{1} {-}{ }Z_{w, 2})   \nonumber \\[-3mm]
& \; \quad  + (\mathbbm{1} {+} Z_{v, 1}) (\mathbbm{1} {+} Z_{v, 2})  (\mathbbm{1} {+} Z_{w, 1}) (\mathbbm{1} {+} Z_{w, 2}) \nonumber \\[1mm]
& \; \quad + (\mathbbm{1} {+} Z_{v, 1}) (\mathbbm{1} {-} Z_{v, 2})  (\mathbbm{1} {+} Z_{w, 1}) (\mathbbm{1} {-} Z_{w, 2}) \nonumber \\
& \; \quad + (\mathbbm{1} {-} Z_{v, 1}) (\mathbbm{1} {+} Z_{v, 2})  (\mathbbm{1} {-} Z_{w, 1}) (\mathbbm{1} {+} Z_{w, 2}) \Big)  \nonumber \\
\phantom{H} {=}  \sum_{v, w=1}^n & 4A_{vw} \Big( Z_{v,1}Z_{v,2}Z_{w,1}Z_{w,2} {+}  Z_{v,1}Z_{w,1} {+}  Z_{v,2}Z_{w,2}\Big) \nonumber \\
&+ c_1 \mathbbm{1}, \label{eq:4-colors}
\end{align}
where the irrelevant constant term in the last line can be omitted.

Analogously, in the case of $k=2^m$ colors, we can label the possible colors by $m$ bits. To each vertex $v$ of the graph, we assign a string of $m$ bits $b_{v,j}$ ($j=1, \ldots, m$) which labels the color of $v$. Considering the usual correspondence between $b_{v,j}$ and $(\mathbbm{1}-Z_{v,j})$, for a graph with adjacency matrix $A$, the graph coloring problem can be embedded into the ground state problem  
of the $(n \log k)$-qubit Hamiltonian  
\begin{align}
H {=} \sum_{v, w =1}^n A_{vw}\sum_{ \underline{a} \in \{0,1 \}^m} \prod_{\ell=1}^m (\mathbbm{1} {+} (-1)^{a_\ell} Z_{v, \ell}) (\mathbbm{1} {+} (-1)^{a_\ell} Z_{w, \ell}), \label{eq:k-color}
\end{align}
since the computational basis state $\otimes_{v,j}\lvert b_{v,j}\rangle$ is a ground state (in this case a $0$-energy state) of $H$ iff the bitstrings $b_{v,j}$ provide a proper coloring of the graph.

If the number of colors $k$ is not a power of $2$, i.e., $ 2^{m-1} < k < 2^{m}$ , then we again label the colors with $m$-long  bitstrings $(b_1,b_2, \ldots b_m)$, but only those are allowed for which
$ \sum_{j=1}^m 2^{k-j}b_j < k$ is satisfied. One can consistently add new terms to the Hamiltonian 
such that the non-allowed bit-strings are  penalized, as we show in \cite{Adam}. 
In particular, for the case of $k=3$ colors, we will have a Hamiltonian that is a sum of two terms: the first being the same Hamiltonian as for the problems of 4 colors, Eq.~\eqref{eq:4-colors}, and the second term being $\sum_{v=1}^{n} (1-Z_{v,1})(1-Z_{v,2})$ that penalizes the non-allowed $(b_{v,1}, b_{v,2}) =(1,1)$ assignments, while giving zero for the allowed $(b_{v,1}, b_{v,2})$ values.

\section{Quantum Annealer Experiments}
\label{sec:annealer}
% Graph coloring experiment on the D-Wave Quantum Annealer}

In this section, to explore current possibilities of quantum hardware, we present a study to uncover the main limiting factors of graph coloring solved with quantum annealer (QA) devices. We created a series of experiments to be performed with the currently available D-Wave QA hardware.

\subsection{Quantum Annealing for the Coloring Problem}
% \subsubsection{Quantum Annealing heuristics for QUBO problems}
When graph coloring is reformulated as a QUBO problem, as discussed in Sec.~\ref{sec:gc_as_qubo}, its cost function is equivalent to an Ising model and its global optimum can be approximated by QA \cite{kadowaki1998quantum, farhi2002quantum}.  We test the limit of this approach using the commercially available quantum annealer, the D-Wave 2000Q which implements a programmable Ising spin network using superconducting flux qubits \cite{johnson2011quantum}. %\cite{johnson2011quantum}. This QPU implements a programmable Ising spin network using superconducting flux qubits \cite{johnson2011quantum}.

The D-Wave 2000Q QPU quantum annealer has at most $2048$ available qubits, and has a C$16$ Chimera topology (the \emph{working graph}) consisting of a $16 \times 16$ matrix of $8$-qubit bipartite graphs. 
To create a bridge between logical and physical representation of qubits, a technique called minor-embedding maps logical qubits to physical ones.
% with a heuristic search which can be automated. 

Since minor-embedding is an NP-hard problem, heuristic algorithms can be employed to find an embedding of the coloring problems \cite{cai2014practical}.
While theoretically any graph with $n$ nodes can be minor-embedded into a Chimera graph with $O(n^2)$ nodes \cite{bienstock1994algorithmic}, several studies of minor-embedding algorithms suggest that, the set of completely embeddable problems is also limited by the effectiveness of the minor-embedding algorithm \cite{boothby2016fast, yang2016graph, rieffel2015case}.

Another problem with the minor embedding is that it connects physical qubits otherwise not connected (due to the sparse Chimera structure), creating \emph{chains} of physical qubits that tend to grow very long in case of large problem complexity. 
We measured how the lengths of these physical qubit chains, created by automated minor-embedding, are effected by the number of nodes and colors and the edge probability of the ER graphs affect the lengths of these physical qubit chains, created by automated minor-embedding. 
Fig.~\ref{fig:embed} summarizes the successful embeddings for random graphs, as the result of more than $50 000$ embedding trials. 

\begin{figure}[t!]
\hspace*{-2mm}
% \centerline{\resizebox{0.36\textwidth}{!}{\input{num_nodes_num_colors_prob_edge__max_chain_len.pgf}}}
\centering
{\includegraphics[width=0.36\textwidth]{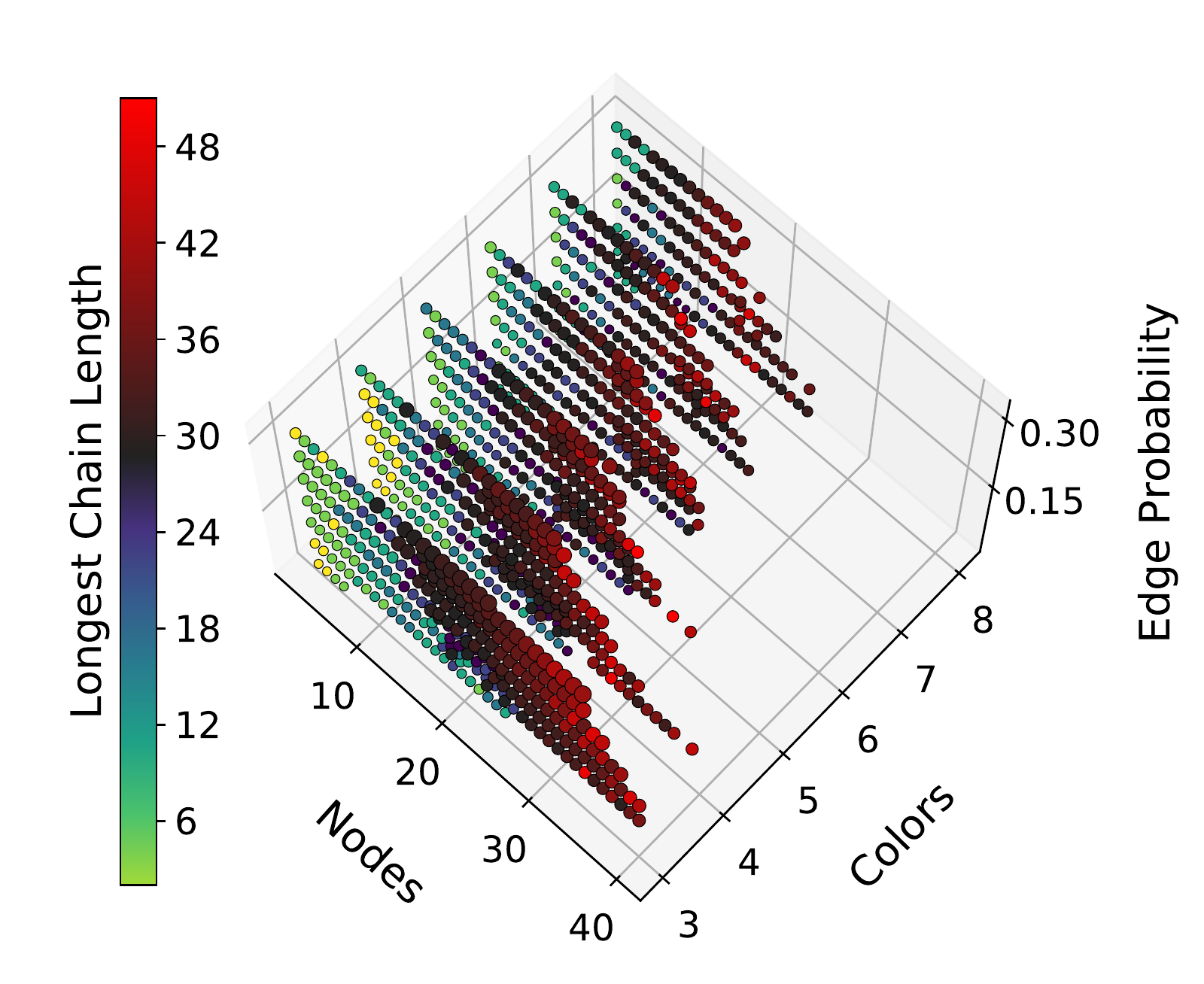}}
\caption{Longest chain lengths, indicated by colors, as the problem complexity increases. Results of more than $50 000$ attempts to minor-embed problems into the Chimera structure of the D-Wave 2000Q QPU are presented; the included lengths are the longest ones of the best embeddings found.}
\label{fig:embed}
\end{figure}

% While in our research, we focused on the performance of the quantum annealer, we also include measurements of embedding results as function of problem parameters. 
% In our statistics, we tried to embed problems of different parameters into the Chimera structure of the 2000Q QPU. 
% Fig.~\ref{fig:embed} shows how chain lengths were affected by the growth of problem complexity.

% Our findings are in concert with these results.
% We should note, however, that in future generations of quantum annealers, the connectivity of the working graph will likely increase significantly, reducing the negative effect of minor-embedding.

% When a problem has constraints other than operating on binary variables, $Q$ also includes additive penalty terms, guiding the annealing process towards acceptable solutions.
% A general way of solving hard combinatorial problems on a D-Wave machine is to convert the problem into the  QUBO form, which acts as a mathematical framework of problem formulation.
% Once the problem has the right form, it can be solved by the annealer, after which it should be transformed back into the original problem domain.

% In order to use quantum annealing to solve a $k$-coloring problem, one needs to formulate to problem as either an Ising spin model or a Binary Quadratic Optimization (QUBO) model \cite{glover2018tutorial}. 
% The Hamiltonian in \eqref{eq:color_ising} can be converted using D-Wave Systems' software stack.

 \subsection{Coloring Experiments with Random Graphs}

\begin{table}[t!]
% \caption{Experiment dimensions}
% \caption{Coloring problem dimensions}
\caption{Summary of D-Wave QA minor-embeddings and colorings for the Erd\H{o}s-R\'{e}nyi random graphs.}
\begin{center}

\begin{tabular}{|l|r|}
\hline
Colors & $3 - 8$ \\
\hline
Graph size & $3 - 40$ \\
\hline
Edge probability [\%] & $2 - 32$ \\
\hline
Max. problem volume & $96.0$ \\
\hline
Connected graphs & $3837$ \\
\hline
% Embedding success rate  & $54.29\%$ \\
% \hline
Successful embeddings & $1677$ \\
\hline
Max. embedded problem volume & $40.96$  \\
\hline
Coloring success [\%] & $27.46$ \\
\hline
\end{tabular}

% \begin{tabular}{|l|c|}
% \hline
% \thead{Number of nodes} & \makecell{$[3, 40]$} \\
% \hline
% \thead{Number of color} & \makecell{$[3, 8]$} \\
% \hline
% \thead{Graph edge probability} & \makecell{$[0.02, 0.32]$} \\
% \hline
% \thead{Number of graphs} & \makecell{$1678$} \\
% \hline
% \thead{Embedding success rate}  & \makecell{$54.29\%$} \\
% \hline
% \thead{Number of embedded problems} & \makecell{$834$} \\
% \hline
% \thead{Coloring success rate} & \makecell{$27.46\%$} \\
% \hline
% \end{tabular}

\label{tab:color_succ}
\end{center}
\end{table}

%{\color{red}We used random graphs, based on the \emph{Erd\H{o}s-R\'{e}nyi model}, to test the quantum annealing algorithm for the $k$-coloring problem. The Erd\H{o}s-R\'{e}nyi model \cite{erdos1959random} is the most well-known family of random graph ensembles, usually denoted by $G(n, p)$, that prescribes random generation of graphs with $n$ nodes, each node having $p$ probability of being connected by an edge.} 

To test the quantum annealing algorithm for the $k$-coloring problem, we used Erd\H{o}s-R\'{e}nyi random graphs generated according to the $G(n,p)$ model, i.e., a graph of $n$ nodes is generated by randomly including each edge with probability $p$.
The \textit{average connectivity} $c$ of such graphs is given by  $ c= pn$. 
%To measure the validity of a coloring solution, one needs to know the \emph{chromatic number} of the graph in question. The chromatic number ($\chi(G)$) of a graph tells us the theoretical lower limit of the number of colors needed to color the graph. Finding this number in case of random graph is NP-complete \cite{gary1979computers}.

% Although it was shown, that in case $G$ is a planar graph, then $\chi(G) = 4$ \cite{appel1977}, in general, finding this number in case of random graph is far from trivial, and is an active area of research, in fact, finding chromatic number is NP-complete \cite{gary1979computers}.

It is known that there exists a threshold of average graph connectivity above which the graph cannot be colored with $k$ colors, the threshold grows asymptotically as $2k \ln k$ \cite{luczak1991chromatic}.
For smaller number of colors there are different ways to estimate this threshold. For example, it was shown by a heuristic local search algorithm \cite{Mulet_2002}, that is based on Potts spin glass model representation of graph coloring, that one can find $3$, $4$, and $5$-colorings of random graphs with at most $4.69$, $8.9$, $13.69$ average connectivity, respectively. 
Our dataset consisted of quantum anneals with more connected graphs that could be optimally colored, however, we paid attention to these limits in our evaluation.

\subsection{Experimental Results and Findings}
% \subsection{Experimental Performance Characteristics}

To measure the problem size, we use the term \textit{problem volume}, which is simply the multiplication of the problem parameters ($v = p  n  k$). 
On small scales ($v < 7$), the D-Wave annealer was able to solve every single problem, including the evaluation graph coloring problems mentioned in Sec.~\ref{sec:QAOA}. 
As the logical connectivity of the problem graph 
% increases
increased, the quality of the solutions started to degrade, indicated by slowly arising coloring errors. 
These errors formed either as missing node colors, or as pairs of adjacent nodes sharing the same colors. 
The summation of these errors is shown in Fig.~\ref{fig:solerr}.

\begin{figure}[t!]
\centering

% \subfloat[Correlation between problem volume and chain length.\label{subfig:clen_numq}]{
% \resizebox{0.36\textwidth}{!}{\input{complexity_avg_chain_len__num_errors.pgf}}}

% \subfloat[Coloring quality as the number qubits increases.\label{subfig:phy_numq}]{
% \resizebox{0.36\textwidth}{!}{\input{num_qubits_num_phy_qubits__num_errors.pgf}}}

\subfloat[Correlation between problem volume and chain length.\label{subfig:clen_numq}]{
{\includegraphics[width=0.36\textwidth]{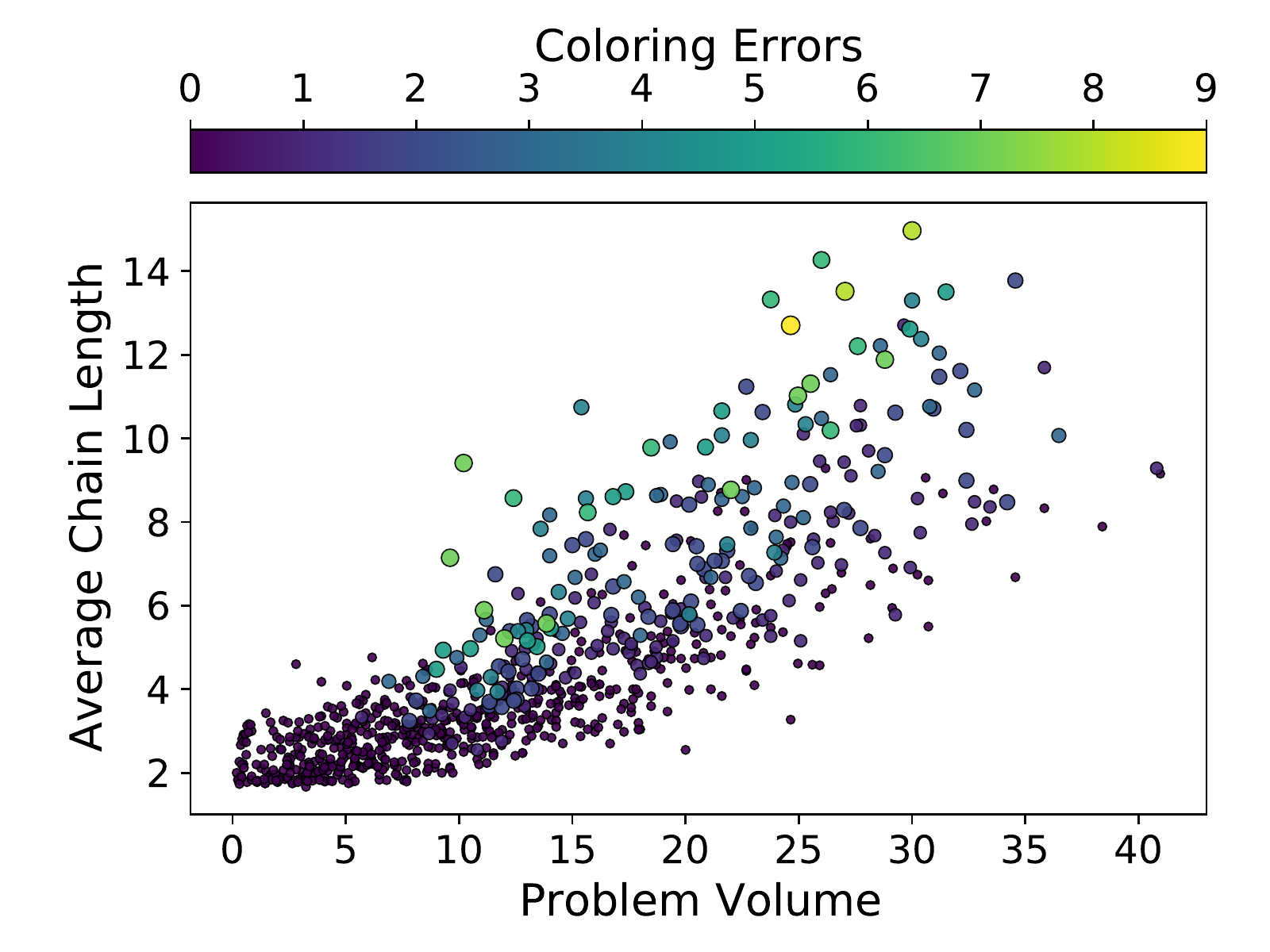}}}

\subfloat[Coloring quality as the number of physical and logical qubits increases.\label{subfig:phy_numq}]{{\includegraphics[width=0.36\textwidth]{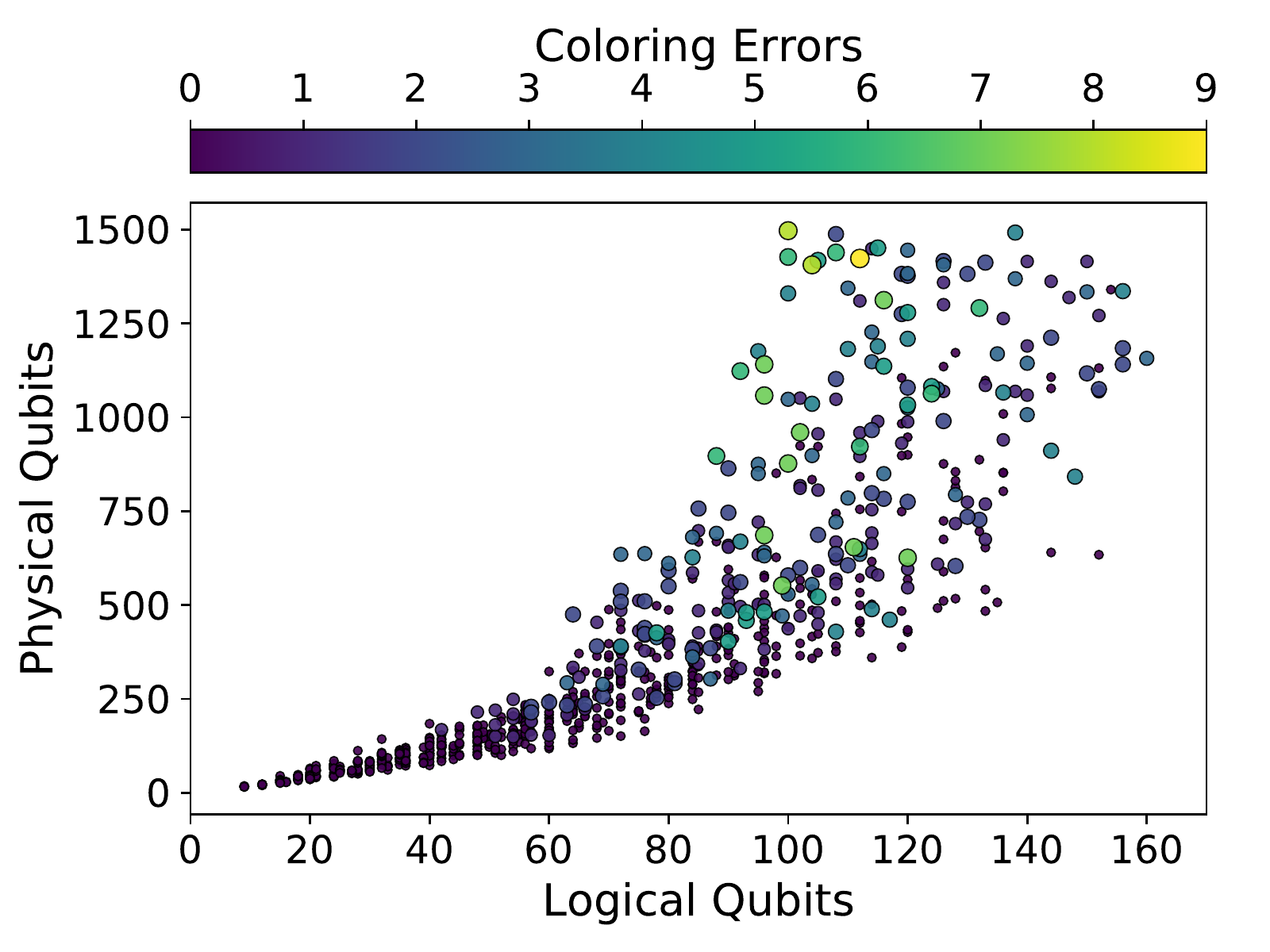}}}
\caption{
The quality of the D-Wave QPU solutions for  graph coloring problems described in Sec.~\ref{sec:QAOA}. More than $200$ random coloring problems of different volume were selected. The number of coloring errors (missing/adjacent colors) is indicated by the color and size of the scatter points. 
Sub-figure~\protect\subref{subfig:clen_numq} depicts how coloring errors arise as the maximal chain lengths of the minor-embedded problem increases with the problem volume, whereas \protect\subref{subfig:phy_numq} shows the physical qubits required for different problem sizes (measured in logical qubits), and how the number coloring errors grow with them.}
\label{fig:solerr}
\end{figure}

While the D-Wave machine performed well on the smaller problems, the illustrations show how it failed to solve most of the complex problems due to the sparse connectivity of the working graph. 
However, we believe, that by using custom embedding procedures, or by fine-tuning the parameters of the solver, these results could improve significantly \cite{yarkoni2019boosting}. 
The biggest random graph problem ($27$ nodes) that we could solve on D-Wave is depicted in Fig.~\ref{fig:big_graph}.

\begin{figure}[t!]
\hspace*{-2mm}
\centering
{\includegraphics[width=0.36\textwidth]{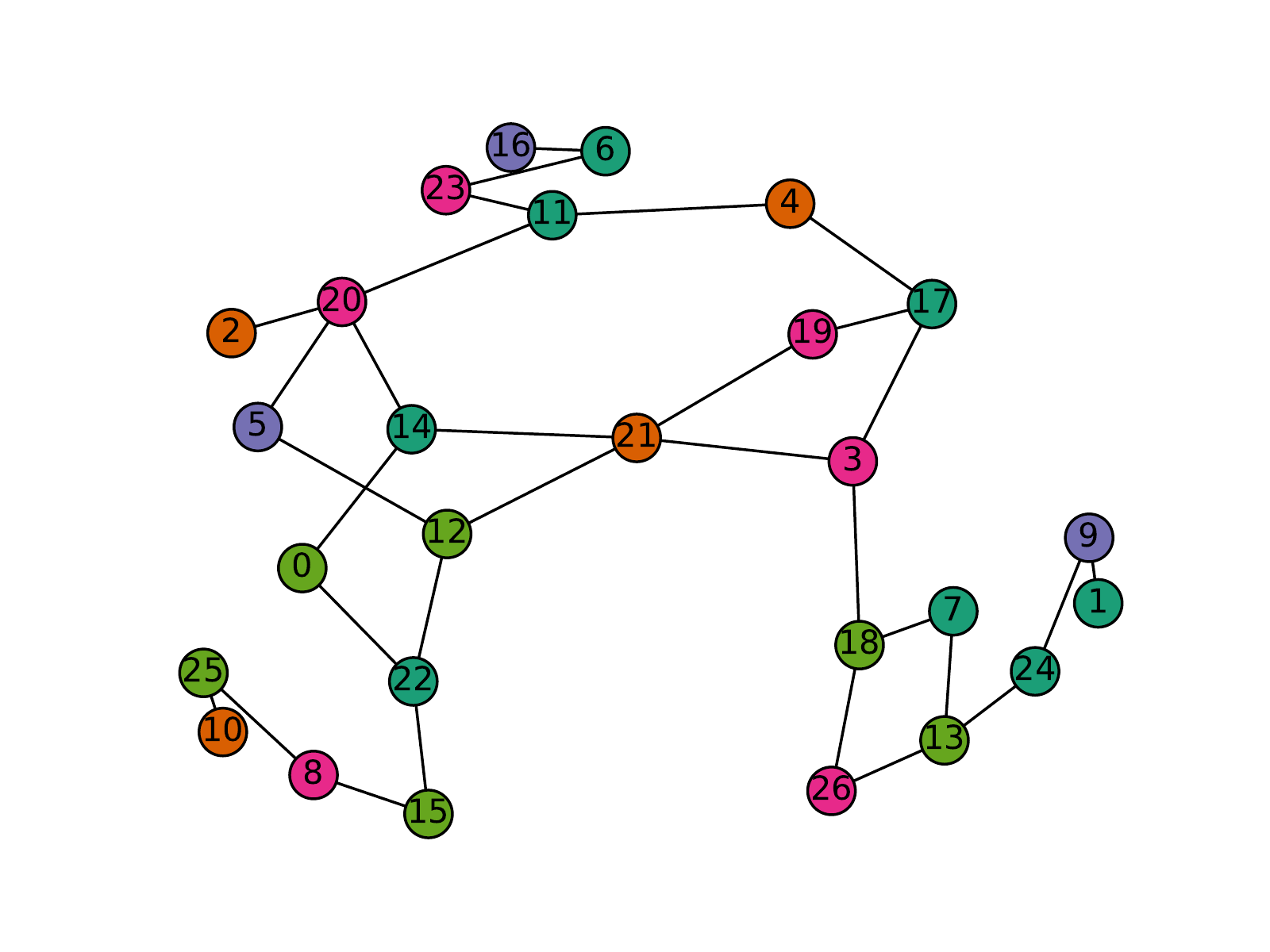}}
\caption{A $27$-node random graph with $0.1$ edge probability, successfully colored with with $5$ colors by the D-Wave quantum annealer (with problem volume of $13.5$). This represents the highest complexity random graph coloring problem that could be solved without fine tuning the parameters of the QPU.
}
\label{fig:big_graph}
\end{figure}

It is worth mentioning, that we were able to solve more than $90\%$ of the embeddable problems perfectly with simple Tabu-search algorithm, with the search run-times restricted to a couple of seconds. 
For this purpose, we used the algorithms provided by D-Waves' Hybrid framework. 

We also ran the problems on the D-Wave Leap's Hybrid Solver, which is a cloud-based quantum-classical solver. 
The hybrid solver managed to solve all (except one) of the hardest problems (ranked by the embedding chain lengths) with computing time limited to $3$ seconds (which is the minimum time limit, for this solver). 
It should be noted, however, that the QPU time per run used by the hybrid solver exceeded twice the time that was required for a pure QPU sampling run.

\section{Simulations of the standard and space-efficient QAOA for graph coloring}
\label{sec:QAOA}
In this section, we present results for the simulation of the newly introduced space-efficient QAOA algorithm for the coloring problem and compare it to the standard approach that uses the QUBO embedding. We refer to the evaluation graphs shown in Fig.~\ref{fig:graphs} with the following notation: $n-k$, where the $n$ represents the number of graph nodes and $k$ denotes the number of colors.

%This algorithm is considered to be one of the most promising approaches towards using near-term quantum computers for practical application, and its experimental feasibility has been recently demonstrated by a novel work \cite{arute2020quantum}.

\subsection{Quantum Approximate Optimization Algorithm}

The Quantum Approximate Optimization Algorithm, introduced originally in Ref.~\cite{farhi2014quantum}, is considered to be  one of the most promising approaches towards using near-term quantum computers for practical application. Its experimental feasibility has been recently demonstrated by a current Google experiment \cite{arute2020quantum}.

The purpose of QAOA is to find an approximate solution ground-state energy of a classical cost
Hamiltonian ${{H}_{c}}$, which usually encodes some combinatorial problem. 
The algorithm starts with applying Hadamard gates on all qubits, i.e., the state of the system is initially transformed to $\left| + \right\rangle^{\otimes N} $.  Next, we apply sequentially the cost Hamiltonian ${{H}_{c}}$ and a mixer Hamiltonian ${H}_{m}$ for time-parameters $\beta_i$ and $\gamma_i$, respectively. After $p$ iterations, the resulting state is of the form:
$U({{H}_{m}},{{\gamma }_{p}})U({{H}_{c}},{{\beta }_{p}})\cdots U({{H}_{m}},{{\gamma }_{1}})U({{H}_{c}},{{\beta }_{1}})\left| + \right\rangle^{\otimes N} $. One then extracts the expectation value of $H_c$ in the given state and using some classical optimizer updates the parameters $\beta_i$ and $\gamma_i$ until the minimum value for a given level-p. When building the quantum circuit for a QAOA, one has to decompose $U({{H}_{c}},{\beta })$ into single rotation gates and CNOTs. In the standard case $H_c$ contains only single-body and two-body terms, $Z_i$ and $Z_i Z_j$. Our space-efficient method uses higher order Hamiltonians. In particular, in our simulations we will have Hamiltonians of the form of Eq.~\eqref{eq:4-colors}, containing fourth-order terms, which can be decomposed as shown in Fig.~\ref{fig:decomp}.

A large amount of studies has been performed to characterize properties of QAOA algorithms, in general and for different application types. These include, both rigorous proofs of computational power and reachability properties \cite{morales2020universality, lloyd2018quantum, hastings2019classical, farhi2016quantum, farhi2020quantum} as well as characterization through heuristics and numerical experiments and extensions of the algorithm \cite{hadfield2019quantum, do2020planning, akshay2020reachability, garcia2019quantum, ho2019efficient, yao2020policy, matos2020quantifying, zhu2020adaptive, wierichs2020avoiding}.

\begin{figure}[t!]
\centering
\subfloat[Graph $A$ \newline $4-3$\label{subfig:graphAB}]{\includegraphics[width=0.13\textwidth, trim={6cm 2.5cm 6cm 2cm}, clip]{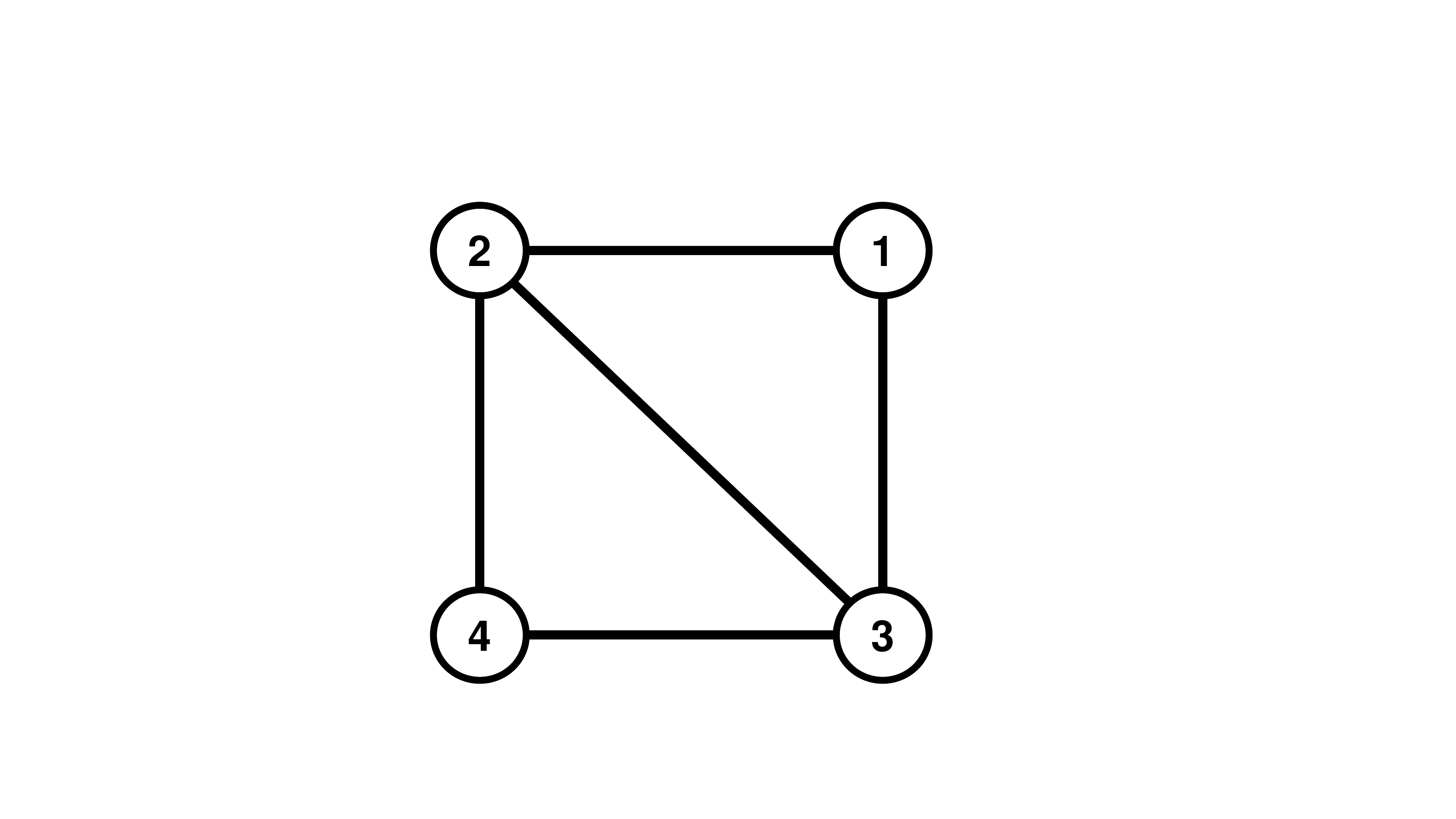}}
\subfloat[Graph $B$ \newline $5-4$\label{subfig:graphF}]{\includegraphics[width=0.13\textwidth, trim={6cm 2.5cm 6cm 2cm}, clip]{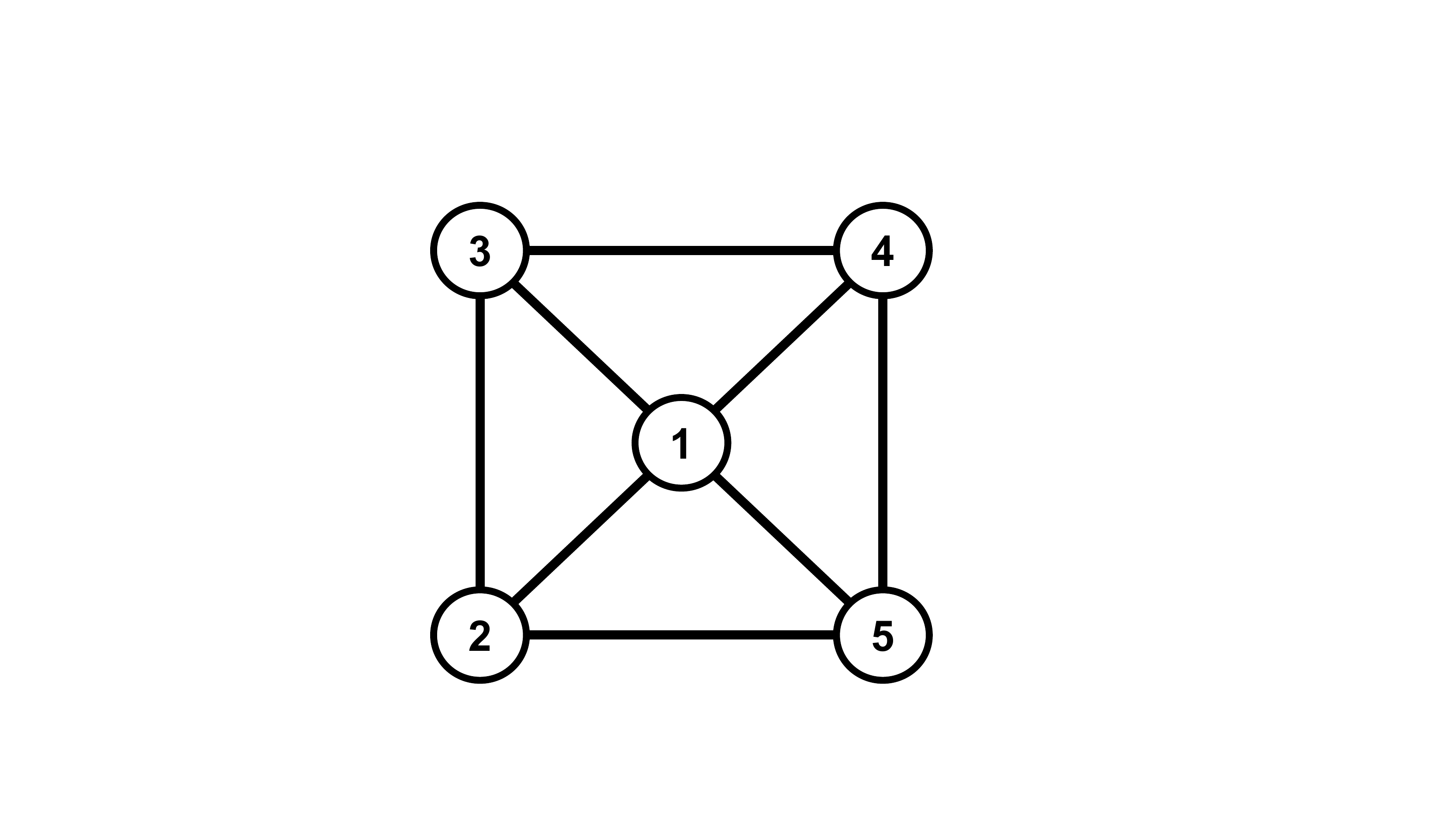}}
\subfloat[Graph $C$ \newline $6-4$\label{subfig:graphH}]{\includegraphics[width=0.13\textwidth, trim={6cm 2.5cm 6cm 2cm}, clip]{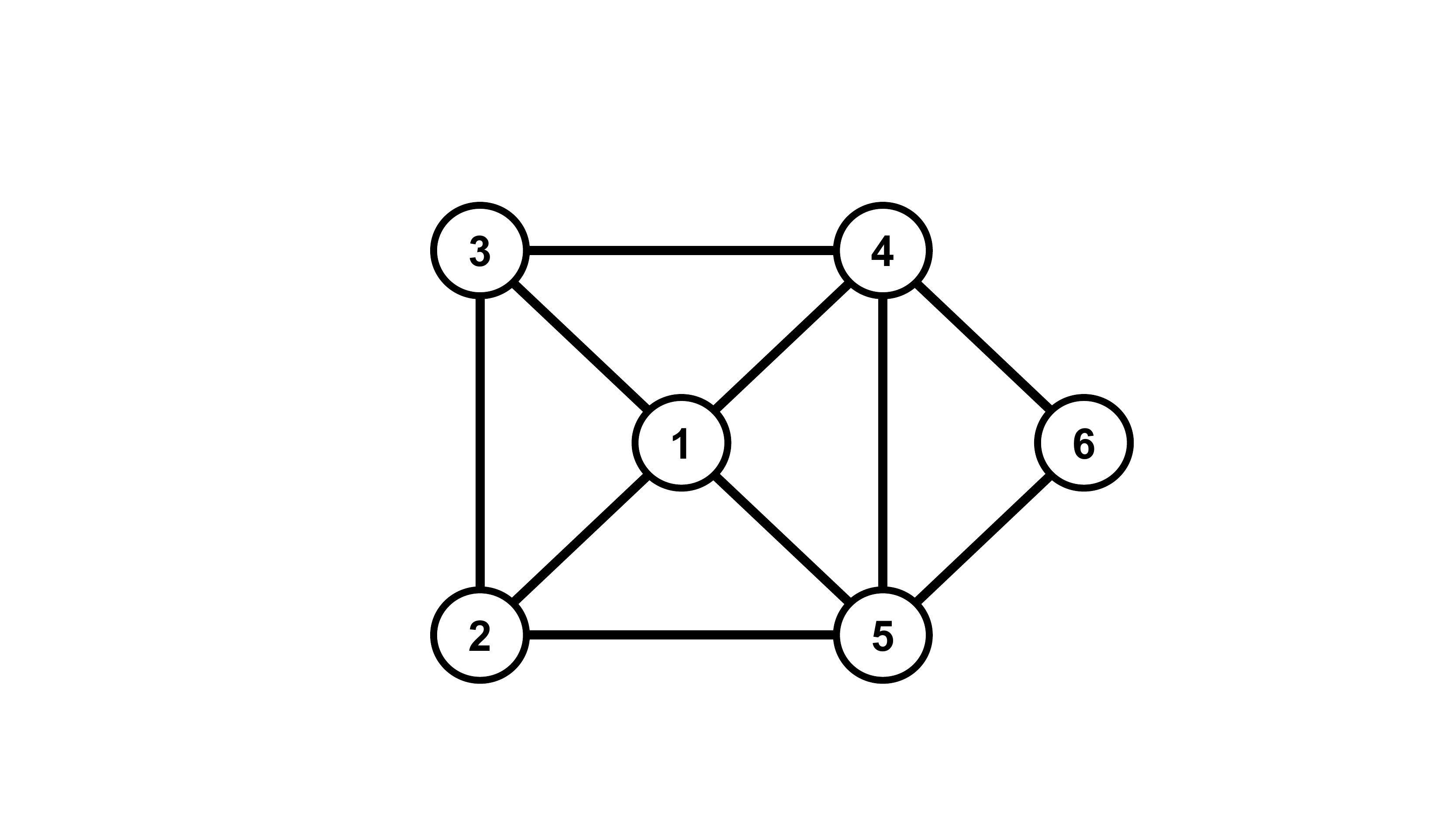}}
\caption{Graph coloring problems used for the performance tests of the space-efficient QAOA method. Graph $A$ with $4$ nodes is used for a $3$-color problem, Graph $B$ contains $5$ nodes and colored with $4$ colors, while Graph $C$ has $6$ nodes and to be colored with $4$ colors. The $n\cdot k$ is twice as big for Graph $C$, than for Graph $A$. The space-efficient method reduces the problem representation size exponentially in the number of colors.}
\label{fig:graphs}
\end{figure}

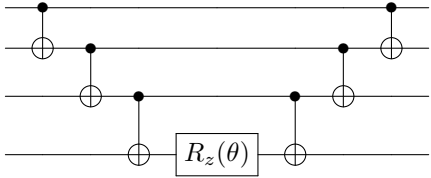
\begin{figure}[t!]%[htp!]
\[
\Qcircuit @C=1em @R=1em  {
& \ctrl{1} & \qw & \qw & \qw & \qw & \qw & \ctrl{1} & \qw \\
& \targ & \ctrl{1} &\qw & \qw & \qw & \ctrl{1} &  \targ & \qw \\
& \qw & \targ & \ctrl{1} & \qw & \ctrl{1} & \targ &  \qw & \qw \\
& \qw & \qw  & \targ & \gate{R_z(\theta)} & \targ & \qw &  \qw & \qw \\
}
\]
\caption{Decomposition of $\textrm{e}^{i \theta( Z \otimes Z \otimes 
Z \otimes Z)/2}$. The task of gate sequence decomposition of the exponentiated Hamiltonian terms similar to Eq.~\ref{eq:4-colors} can be achieved by the symmetric circuit of CNOTs and a central Z-rotation gate. } \label{fig:decomp}
\end{figure}

%Sequential linear-quadratic programming 
Finally, let us mention that an important aspect of running a QAOA-based method is the choice of classical optimizer method. The Nelder-Mead algorithm \cite{nelder_simplex_1965} and constrained optimization by linear approximation (COBYLA) \cite{powell_direct_1998} are commonly used algorithms since they require a low number of function evaluations. Other approaches, such as the sequential least squares programming (SLSQP) \cite{kraft1988software} can be efficient if the search space of the problem is bigger. These methods can be found in SciPy Python package, and Qiskit Aqua has already integrated them into its sub-modules, enabling us to automatically optimize variational quantum circuits.  We chose to use the $NestrovMomentum$ \cite{ruder_overview_2016} variation of the Gradient Descent method. This optimizer shows a good performance in shallow and deep circuits when the momentum and learning rate values are set carefully. We used the PennyLane quantum machine learning framework \cite{bergholm2018pennylane} combined with TensorFlow.
We found that in deep QAOA circuits the backpropagation method provided by TensorFlow  showed a better convergence rate than the Parameter-shift rule, while in shallow circuits both methods have the same performance.

\subsection{Comparison of Standard and Space-Efficient QAOA}
\label{subsec:caseA}

Let us now compare the results for a coloring problem first solved by \emph{i) a standard QAOA approach} presented in Sec.~\ref{sec:qubo-problem} and then by \emph{ii) a space-efficient QAOA} with the method presented in Sec.~\ref{sec:space-efficient-method}. The considered problem is the $3$-coloring of Graph A, shown in Fig.~\ref{subfig:graphAB}. 

Looking at the convergence characteristics, the first problem was solved with a level-$10$ QAOA algorithm, using a circuit of $12$ qubits, resulting in a circuit depth of $170$. The convergence shown in Fig.~\ref{fig:AB} is towards the zero energy level, as we the shifted minimum energy eigenvalue to this value. For the coloring problem, when the gap reaches a size less than $0.75$, the corresponding solution gives a correct coloring of the graph with probability $0.25$. For the original algorithm, this energy level is reached at iteration no.~$240$, while the enhanced algorithm of space-efficient QAOA reaches the same level already at iteration no.~$44$ - showing a substantial improvement in the performance.
Another metric to characterize the performance gain of the new algorithm is the overall CPU time usage. In this metric, the enhancement is showing an improvement of increasing execution speed by $75$ times. Moreover, the new method decreased the memory usage by a factor of  $3.47$.

Taking the probability distribution over the best circuit solution, we can calculate the probability of measuring a state corresponding to a good coloring solution. 
We find the probabilities to be $0.557$ for the original circuit, and $0.883$ for our space-efficient algorithm.

The depth of a single QAOA level for the enhanced circuit is $37$, whereas for the original circuit this number is $17$. However, the enhanced algorithm needs only a level-$6$ QAOA, using a circuit of $8$ qubits. Hence, although the total depth is increased, this enlargement is not that substantial and the number of required qubits and the needed iterations  is far less.

\begin{figure}[t!]
\hspace*{-2mm}
\centering
{\includegraphics[width=0.36\textwidth]{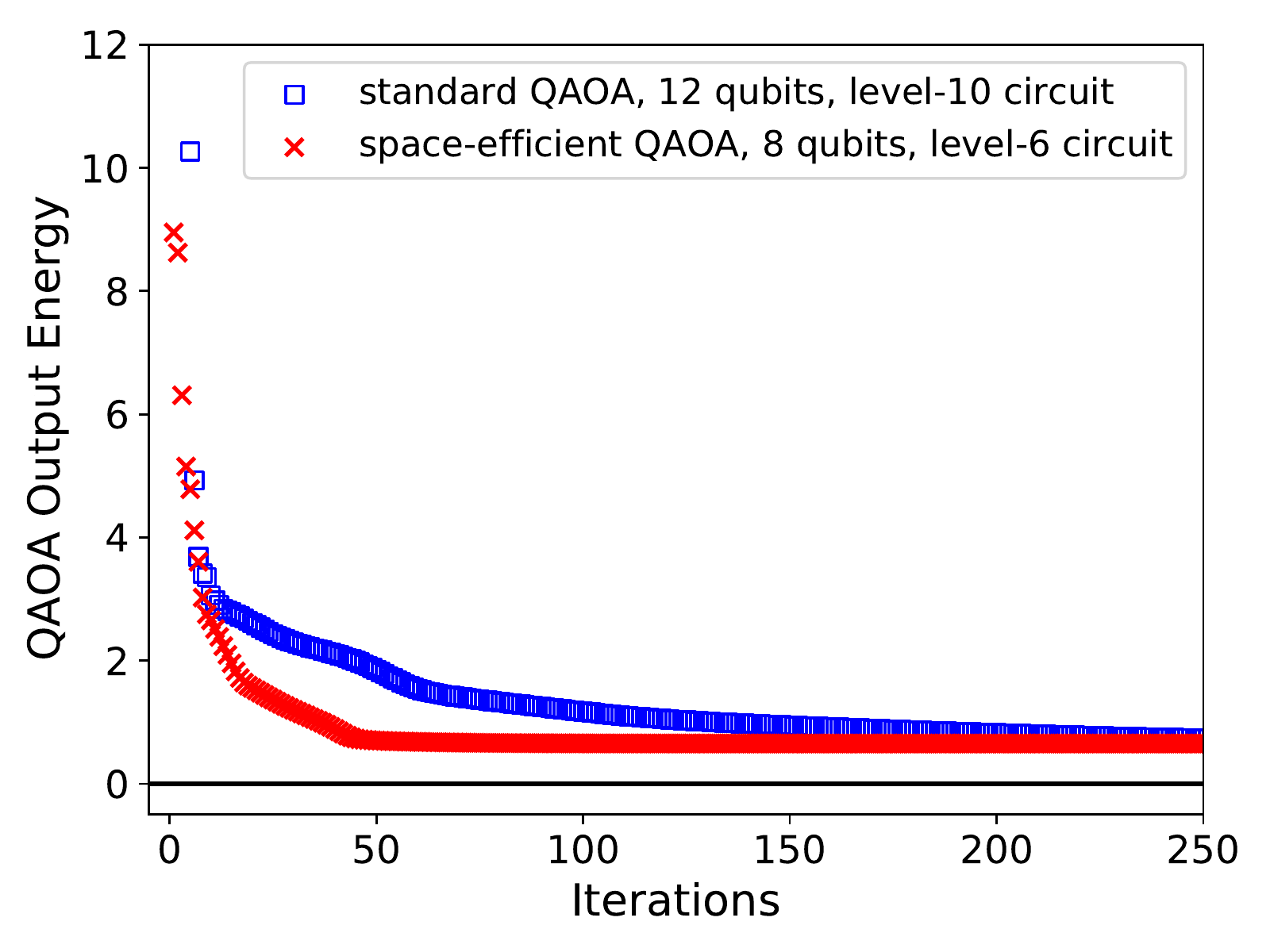}}
\caption{The convergence of the standard (blue square) and the space-efficient (red cross) QAOA simulations for the coloring of Graph $A$. The standard method requires $12$ qubits and level-$10$ circuits to converge, while for the space-efficient method it is sufficient to use $8$ qubits and a level-$6$ circuit. The performance gain from the space-efficient modification is measurable in both the lower number of iterations for reaching near optimum and the overall reduction in CPU time used. 
}
\label{fig:AB}
\end{figure}

\subsection{Application of Stochastic Gradient Descent}
We also studied the convergence properties of both the gradient descent with exact Hamiltonian expectation values and the {\it stochastic gradient descent} (SGD) \cite{sweke2019stochastic, elsafty2020stochastic} obtained from estimating the expectation value from only a finite $n$ number of shots. 

For this investigation, we considered the coloring problem of Graph B represented in Fig.~\ref{subfig:graphF} with  $k=4$ colors.
During the simulations, we used a level-$6$ QAOA, which has a circuit depth of $354$. We found that after the optimization, the probability of measuring a bitstring representing a good coloring of Graph B is $0.845$. 

As shown in Fig.~\ref{fig:F_layer_6}, the convergence of the true gradient descent optimization approaches the global minimum value with an energy distance of $0.75$ already at the iteration no. $13$, while the SGD with $50$ and $100$ shots at iterations $10$ and $17$, respectively.  The global minimum is approached with a distance of $0.5$ after $32$ iterations when using the true gradient descent method and at iteration steps $11$ and $43$ by using SGD with $50$ and $150$ number of shots, respectively. 
That is, the SGD performed on par with the gradient descent using exact expectation values, and it would require only  $150$ measurements per iterations much less than one .

\begin{figure}[t!]
\hspace*{-2mm}
% \centerline{\resizebox{0.36\textwidth}{!}{\input{layer_6.pgf}}}
\centering
{\includegraphics[width=0.36\textwidth]{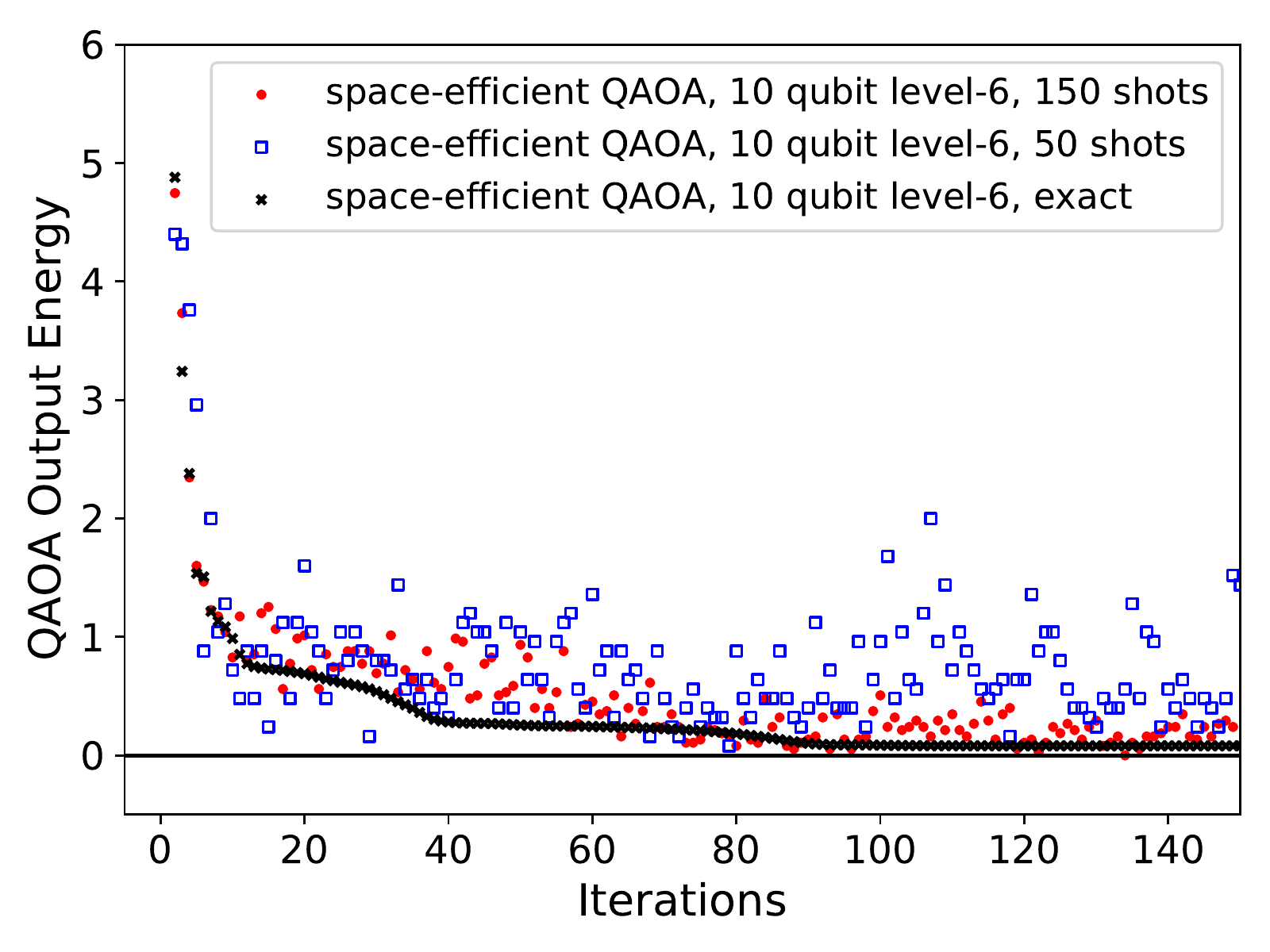}}
\caption{Space-efficient QAOA simulations for the coloring problem of Graph $B$. The figure shows the comparison between the exact expectation value (black cross), and stochastic approximations using $50$ measurements (blue square) and $150$  measurements (red dot). It is clearly visible that the $150$ shot approximation reaches the same level of coloring efficiency as the exact simulation. This method can further decreases the load on a quantum device in terms of the number of runs and measurements required.}
\label{fig:F_layer_6}
\end{figure}

\subsection{Evaluation of the $6$-node, $4$-color problem on Graph $C$}
\label{subsec:caseH}
Here we present our findings on the evaluation problem of Graph C, which is the coloring task described in Fig.~\ref{subfig:graphH}. In this more complex example, we looked for the successful coloring (with $k=4$ colors) of the $6$-node graph. To simulate this problem using the space-efficient embedding technique, it is enough to consider a circuit containing only $12$ qubits. A fast convergence can be achieved by only using a level-$9$ QAOA algorithm which is, once again, lower than the requirements for the simplest problem Graph $A$ solved with the standard embedding. While the convergence pattern is different from the ones seen in Fig.~\ref{subfig:graphAB} and Fig.~\ref{fig:F_layer_6}, we observe significant improvement in convergence characteristics. The QAOA output energy reaches the theoretical minima within $0.1$ gap in only $100$ iterations. Our simulations show that the observed probability of finding the correct solution is as high as $97.4\%$. These results represent a performance improvement of $97.3\%$ as compared to the Graph A simulation in Sec.~\ref{subsec:caseA} in terms of CPU time required.

\begin{figure}[t!]
\hspace*{-2mm}
% \centerline{\resizebox{0.36\textwidth}{!}{\input{H.pgf}}}
\centering
{\includegraphics[width=0.36\textwidth]{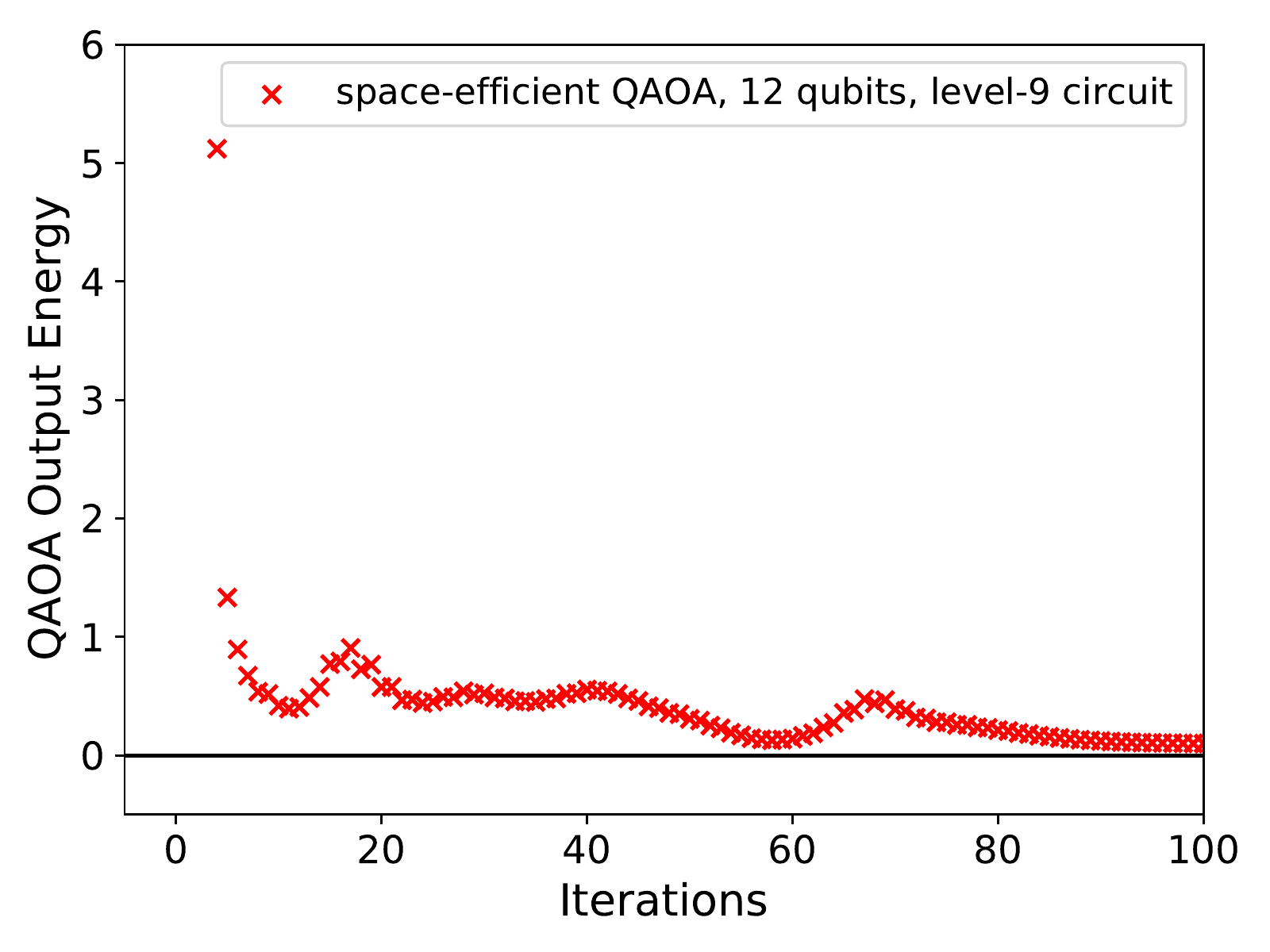}}
\caption{Numerical simulation of the space-efficient QAOA on Graph $C$. A fast convergence can be seen within as low as $100$ iterations by only using a level-$9$ circuit. Although the problem space is of higher complexity than the evaluation problem over Graph $A$, a lower level algorithm performs better with the space-efficient embedding technique.}
\label{fig:H}
\end{figure}

\section{Conclusion and Outlook}
\label{sec:conclusion}
We introduced a space-efficient embedding for quantum circuits solving the graph coloring problem. Through a series of investigations, we presented the performance gain of this method. We showed the limitations of the existing QA hardware solutions and then with various numerical simulations compared the standard and enhanced QAOA circuits. The required circuit width to embed the coloring problem is exponentially reduced in the number of colors; and although the depth of a single QAOA layer is increased, the number of required layers and optimization iteration steps to reach optimal solution are also decreased. 
The presented method and comparative study can be extended to a benchmarking framework for such performance gain analyses. 
Furthermore, analogous space-efficient embedding techniques could be used to improve upon other graph-related quantum optimization methods. We leave this for future work.

\section*{Acknowledgment}

Z.T. and Z.Z. would like to thank the support of the Hungarian Quantum Technology National Excellence Program (Project No. 2017-1.2.1-NKP-2017-00001). Z.Z. acknowledges also support from the Hungarian National Research, Development and Innovation Office (NKFIH) through Grants No. K124351, K124152, K124176 KH129601 and the J\'anos Bolyai Scholarship. A.G. was supported by the Polish National Science Center under the grant agreements 2019/32/T/ST6/00158 and 2019/33/B/ST6/02011.
 \newpage

% \nocite{*}
\bibliographystyle{IEEEtran}
\bibliography{IEEEabrv,qgc}

\end{document}